\documentclass[aps,twocolumn,pra]{revtex4-1}
\usepackage{graphicx} 
\usepackage{amssymb,amsmath,amsthm} 
\usepackage{color}
\providecommand{\openone}{\leavevmode\hbox{\small1\kern-3.8pt\normalsize1}} 
\providecommand{\Prob}{\text{Prob}}

\providecommand{\ket}[1]{|#1\rangle}

\providecommand{\ketbra}[2]{|#1\rangle\kern-2.8pt\langle#2|}
\providecommand{\sg}{\text{\,sg}}
\newtheorem*{theorem*}{Theorem}
\newtheorem*{proposition*}{Proposition}

\definecolor{nred}{rgb}{0.7,0.2,0.2}
\definecolor{nblack}{rgb}{0,0,0}

\def\be{\begin{equation}}
\def\ee{\end{equation}}

\begin{document}
\title{Device-independent witnesses of genuine multipartite entanglement}
\author{Jean-Daniel Bancal$^1$, Nicolas Gisin$^1$, Yeong-Cherng Liang$^1$, Stefano Pironio$^2$ \\[0.5em]
{\it $^1$Group of Applied Physics, University of Geneva, Switzerland} \\
{\it $^2$Laboratoire d'Information Quantique, Universit\'e Libre de Bruxelles, Belgium}}
\date{\today}

\begin{abstract}
We consider the problem of determining whether genuine multipartite entanglement was produced in an experiment, without relying on a characterization of the systems observed or of the measurements performed. We present an $n$-partite inequality that is satisfied by all correlations produced by measurements on biseparable quantum states, but which can be violated by
$n$-partite entangled states, such as GHZ states.  In contrast to traditional entanglement witnesses, the violation of this inequality implies that the state is not biseparable independently of the Hilbert space dimension and of the measured operators. Violation of this inequality does not imply, however, genuine multipartite non-locality. We show more generically how the problem of identifying genuine tripartite entanglement in a device-independent way can be addressed through semidefinite programming.
\end{abstract}

\maketitle

The generation of multipartite entanglement is a central objective in experimental quantum physics. For instance,  entangled states of fourteen ions and six photons have recently been produced \cite{ion,photon}.
In any such experiment, a typical question arises: How can we be sure that \emph{genuine $n$-partite entanglement} was present? A state is said to be genuinely $n$-partite entangled if it is not \emph{biseparable}, that is, if it cannot be prepared by mixing states that are separable with respect to some partition. Consider for instance the tripartite case: a state $\rho_{\mathrm{bs}}$ is said to be biseparable if it admits a decomposition
\be\label{stbis}
\rho_{\mathrm{bs}}=\sum_k \rho_{AB}^k\otimes \rho_{C}^k+\sum_k \rho_{AC}^k\otimes \rho_{B}^k+\sum_k \rho_{BC}^k\otimes \rho_{A}^k\,,
\ee
where the weight of each individual state in the mixture has been included in its normalization; a state that cannot be written as above is genuinely tripartite entangled. Determining whether genuine $n$-partite entanglement was produced in an experiment represents a difficult problem that has attracted much attention recently (see e.g. \cite{review,multent}). The usual approach consists of measuring a witness of genuine multipartite entanglement, or of doing the full tomography of the state followed by a direct analysis of the reconstructed density matrix.

Such approaches, however, not only rely on the observed statistics to conclude about the presence of entanglement, but also require a detailed characterization of the systems observed and of the measurements performed. Consider for instance the following witness of genuine tripartite entanglement:
\be \label{mermin}
M=X_1X_2X_3 - X_1Y_2Y_3- Y_1X_2Y_3 - Y_1Y_2X_3\,,
\ee
where $X_j=\sigma_x$ and $Y_j=\sigma_y$ are the Pauli spin observables in the $x$ and $y$ direction for particle $j$. For any biseparable three-qubit state $\langle M\rangle = \text{tr}\left(M\rho\right)\leq 2$ \cite{addendum}. Thus if we measure three spin-$\frac12$ particles in the $x$ and $y$ direction and find an average value $\langle M\rangle >2$, we can conclude that the state exhibits genuine tripartite entanglement.

Suppose, however, that the measurement $Y_3$ carries a slight (possibly unnoticed) bias towards the $x$ direction. That is, instead of measuring $Y_3=\sigma_y$, we actually measure $Y_3=\cos\theta\sigma_y +\sin\theta\sigma_x$. Then it is not difficult to see, all other measurements being ideal, that the biseparable state
$|\psi\rangle=\frac{1}{2}\left(|00\rangle+e^{-i\phi}|11\rangle\right)_{AB}\otimes\left(|0\rangle+|1\rangle\right)_C$,
where $\phi=\arctan(\sin\theta)$, yields $\langle M\rangle=2\sqrt{1+\sin^2\theta}$ which is strictly larger than 2 for any $\theta\neq 0$. Thus, unless we measure all particles \emph{exactly} along the $x$ and $y$ directions, we can no longer conclude that observing $\langle M\rangle>2$ implies genuine tripartite entanglement.
Importantly, this is not a unique feature of the above witness, but rather all conventional witnesses are, to some extent, susceptible to such systematic errors that are seldom taken into account.

Furthermore, tomography and usual entanglement witnesses typically assume that the dimension of the Hilbert space is known. For instance, in a typical experiment demonstrating, say, entanglement between four ions, we usually view each ion as a two-level system. But an ion is a complex object with many degrees of freedom (position, vibrational modes, internal energy levels, etc). How do we know, given the inevitable imperfections of the experiment, that it is justified to treat the relevant Hilbert space of each ion as two-dimensional and how does this simplification affects our conclusions about the entanglement present in the system \cite{lutk}? Even if it is justified to view each ion as a qubit, is entanglement between four systems really necessary to reproduce the measurement data, or could they be reproduced with fewer entangled systems if qutrits were manipulated instead?

These remarks motivate the introduction of entanglement witnesses that are able to guarantee that a quantum system exhibits ($n$-partite) entanglement, without relying on the types of measurements performed,  the precision involved in their implementation, or on assumptions about the relevant Hilbert space dimension.
We call such witnesses, \emph{device-independent} entanglement witnesses (DIEW). This type of approach was already considered in Refs.~\cite{seevinck,diew}. Note that other solutions to the above problems are possible, such as entanglement witnesses tolerating a certain misalignement in the measurement apparatuses \cite{uff} or the characterization of realistic measurement apparatuses through
squashing maps \cite{lutk}. These types of more specific approaches, however, still require some partial characterization of the system and measurement apparatuses, which is not necessary when using DIEWs. 

Any DIEW is a Bell inequality (i.e. a witness of nonlocality). Indeed, $i)$ the violation of a Bell inequality implies the presence of entanglement, and $ii)$ any measurement data that does not violate any Bell inequality can be reproduced using quantum states that are fully separable~\cite{R.F.Werner:PRA:1989}. The violation of a Bell inequality is thus a necessary and sufficient condition for the detection of entanglement in a device-independent (DI) setting. This observation is the main insight behind DI quantum cryptography \cite{collective,Pironio10}, where the presence of entanglement is the basis of security.

The relation between DIEW for \emph{genuine $n$-partite} entanglement and witnesses of multipartite nonlocality is more subtle. While there exist Bell inequalities that detect genuine $n$-partite nonlocality \cite{svet,collins,svetd}, not every DIEW for $n$-partite entanglement corresponds to such a Bell inequality. Consider for instance, the expression (\ref{mermin}). If no assumptions are made on the type of systems observed and measurements performed, the inequality $\langle M\rangle \leq 2$ corresponds to Mermin's Bell-type inequality \cite{mermin}, i.e., a value $\langle M\rangle >2$ necessarily reveal non-locality, hence entanglement. Moreover, a value $\langle M\rangle >2\sqrt{2}$ guarantees genuine tripartite entanglement \cite{collins,seevinck}. The Mermin expression (\ref{mermin}) can thus be used as a tripartite DIEW. Yet, it cannot be used as a Bell inequality for genuine tripartite non-locality, since a simple model involving communication between two parties only already achieves the algebraic maximum $\langle M\rangle =4$~\cite{collins}.

The objectives of this paper are to formalize the concept of DIEW for genuine multipartite entanglement and initiate a systematic study that goes beyond the early examples given in \cite{seevinck,diew}. Following this line, we start by introducing the notion of quantum biseparable correlations. We then present a simple DIEW for $n$-partite entanglement which is stronger for GHZ (Greenberger-Horne-Zeilinger) states than all the inequalities introduced in \cite{seevinck,diew}.  In the case $n=3$, we also provide a general method for determining whether given correlations reveal genuine tripartite entanglement and apply it to GHZ and W states. Apart from yielding practical criteria for the characterization of entanglement in a multipartite setting, our results also clarify the relation between device-independent multipartite entanglement and mulitpartite nonlocality.

\textbf{1. Biseparable quantum correlations.}
For simplicity of exposition, let us consider an arbitrary tripartite system (the following discussion easily generalizes to the $n$-party case). To characterize in a DI way its entanglement properties, we consider a Bell-type experiment: on each subsystem, one of $m$ possible measurements is performed, yielding one of $d$ possible outcomes. We adopt a black-box description of the experiment and represent the measurements on each of the three subsystems by classical labels $x,y,z\in\{1,\ldots,m\}$ (corresponding, e.g., to the values of macroscopic knobs on the measurement apparatuses) and denote the corresponding classical outcomes $a,b,c\in\{1,\ldots,d\}$. The correlations obtained in the experiment are characterized by the joint probabilities $P(abc|xyz)$ of finding the triple of outcomes $a,b,c$ given the measurement settings $x,y,z$. 

We say that the correlations $P(abc|xyz)$ are \emph{biseparable quantum correlations} if they can be reproduced through local measurements on a biseparable state $\rho_{\mathrm{bs}}$. That is, if there exist a biseparable quantum state \eqref{stbis} in some Hilbert space $\mathcal{H}$, measurement operators $M_{a|x}$, $M_{b|y}$, and $M_{c|z}$ (which without loss of generality we can take to be projections satisfying $M_{a|x} M_{a'|x} = \delta_{a,a'}M_{a|x}$ and $\sum_{a}M_{a|x}=\openone$), such that
\be\label{eq:corrs}
P(abc|xyz) = \text{tr}\left[M_{a|x}\otimes M_{b|y} \otimes M_{c|z}\,\rho_{\mathrm{bs}}\right]\,.
\ee
If given quantum correlations $P(abc|xyz)$ are not biseparable, they necessarily arise from measurements on a genuinely tripartite entangled state, and this conclusion is independent of any assumptions on the type of measurements performed or on the Hilbert space dimension.

Equivalently, biseparable quantum correlations can be defined as those that can be written in the form
\be\label{eq:svetq}\begin{split}
P&(abc|xyz) = \sum_k P^k_Q(ab|xy)P_Q^k(c|z)\\
&+\sum_k P^k_Q(ac|xz)P_Q^k(b|y)+\sum_k P^k_Q(bc|yz)P_Q^k(a|x),
\end{split}
\ee
where $P_Q^k(ab|xy)$ and $P_Q^k(c|z)$ correspond, respectively, to arbitrary two-party and one-party quantum correlations, i.e., they are of the form $P_Q^k(ab|xy)=\text{tr}[M^k_{a|x}\otimes M^k_{b|y}\,\rho^k_{AB}]$ and $P_Q^k(c|z)=\text{tr}[M^k_{c|z}\,\rho^k_C]$ for some unormalized quantum states $\rho^k_{AB}$, $\rho^k_{C}$ and measurement operators $M^k_{a|x}$, $M^k_{b|y}$, $M^k_{c|z}$ (and similarly for the other terms in (\ref{eq:svetq})). Note that here the measurement operators for different bi-partitions do not need to be the same (though this can always be achieved as shown in Appendix \ref{app:sdp}). That any correlations of the form (\ref{eq:corrs}) is of the form (\ref{eq:svetq}) is immediate using the definition (\ref{stbis}) of biseparable states. Conversely, it is easy to see that any correlations of the form (\ref{eq:svetq}) is of the form (\ref{eq:corrs}), see Appendix \ref{app:def}.

\begin{figure}
\includegraphics[width=0.92\columnwidth]{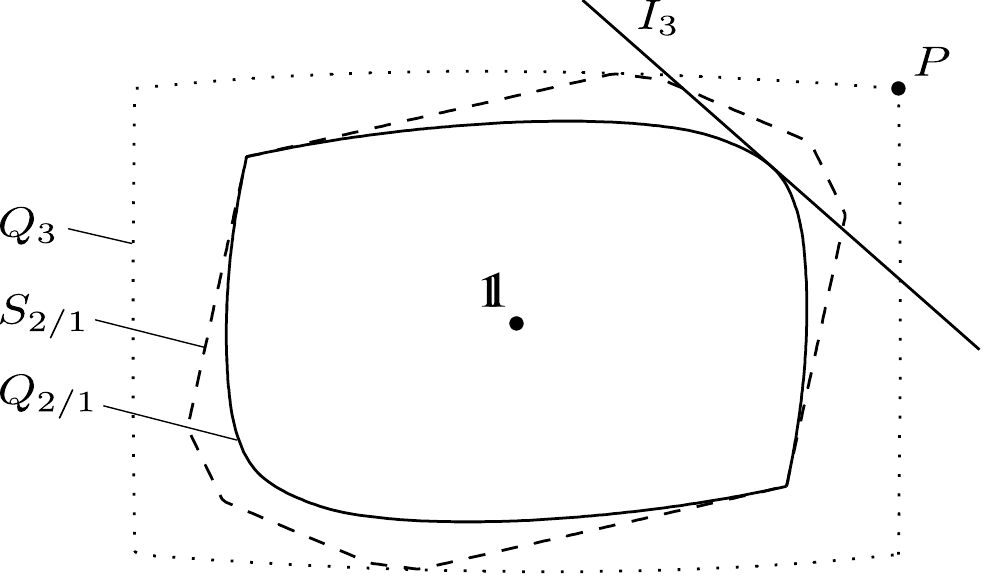}
\caption{A particular slice of the space of tripartite correlations with 3 settings and 2 outcomes representing schematically the sets of general quantum correlations ($Q3$), Svetlichny correlations ($S_{2/1}$) and biseparable quantum correlations ($Q_{2/1}$). The point $\openone$ corresponds to random correlations and $P$ to the GHZ correlations maximally violating the DIEW (\ref{eq:witness}), which is represented by the straight line $I_3$; note that a DIEW can be violated by Svetlichny-local correlations. (The Svetlichny polytope $S_{2/1}$ can be determined exactly using linear programming, while $Q_3$ and $Q_{2/1}$ can be approximated efficiently using SDP techniques, see main text.)}
\label{fig:fig}
\end{figure}

Let $Q_3$ denote the set of tripartite quantum correlations and $Q_{2/1}\subset Q_3$ the set of biseparable quantum correlations. From (\ref{eq:svetq}), it is clear that $Q_{2/1}$ is convex and that its extremal points are of the form $P_{Q}^{ext}(ab|xy) P^{ext}_Q(c|z)$ where $P_Q^{ext}(ab|xy)$ is an extremal point of the set $Q_2$ of bipartite quantum correlations and $P^{ext}_Q(c|z)$ an extremal point of the set $Q_1$ of single-party correlations (the extreme points of $Q_1$ are actually classical, deterministic points). Since the set $Q_{2/1}$ is convex, it can be entirely characterized by linear inequalities. Those linear inequalities separating $Q_{2/1}$ from $Q_3$ correspond to DIEWs for genuine tripartite entanglement. Since $Q_{2/1}$ has an infinite number of extremal points, there exist an infinite number of such inequalities.
Note that the set of local correlations $P(abc|xyz)=\sum_k P^k(a|x)P^k(b|y)P^k(c|z)$ is contained in $Q_{2/1}$. This implies that any DIEW for genuine tripartite entanglement is a Bell inequality (though not necessarily  a tight Bell inequality). Note also that the decomposition (\ref{eq:svetq}) corresponds to a Svetlichny-type decomposition \cite{svet} where all bipartite factors are restricted to be quantum, whereas less restrictive constraints (or even none in Svetlichny original definition $P(abc|xyz)=\sum_k P^k(ab|xy)P^k(c|z)+\ldots$ ) are imposed on these bipartite terms in the definitions of multipartite non-locality \cite{svet}. It follows that the set of genuinely bipartite non-local correlations is larger than the set of biseparable quantum correlations as illustrated in Fig.~\ref{fig:fig}. Thus while any Bell inequality detecting genuine tripartite non-locality is a DIEW for genuine tripartite entanglement, the converse is not necessarily true. 
All these observations extend to the $n$-party case.

\textbf{2. A DIEW for $n$-partite entanglement.}
We now present a DIEW for $n$ parties, where each party $i$ performs a measurement $x_i\in\{1,2,3\}$ and obtains an outcome $a_i\in\{-1,1\}$. We denote $E(\bar{x})$ the correlator associated to the measurement settings $\bar x=(x_1\ldots,x_n)$, i.e. the expectation value $E(\bar{x})=\sum_{\bar{a}} P(\bar{a}|\bar{x})\,\prod_{i=1}^n a_i$, where $\bar a=(a_1\ldots,a_n)$ denotes an $n$-tuple of outcomes.
Let $E_n^k=\sum_{\bar{x}} \delta(\sum_{i=1}^n x_i=k) E(\bar{x})$ be the sum of correlators $E(\bar x)$ for which the measurement settings $x_i$ of the $n$ parties sum up to $k$. Let $f_k$ be a function such that $f_{k+3}=-f_k$ and taking successively the values $[1,1,0]$ on the integers $k=0,1,2$. 
Then the inequality 
\be\label{eq:witness}
I_{n} 	=\sum_{k=n}^{3n} f_{k-n}\,E_n^k\leq 2 \times 3^{n-3/2}
\ee
is satisfied by all biseparable quantum correlations, and is thus a DIEW for genuine $n$-partite entanglement. The proof of this statement is based on the decomposition (\ref{eq:svetq}) and is given in Appendix~\ref{app:diew}. The Svetlichny bound associated to the expression $I_n$,  on the other hand,  is easily found to be $4 \times 3^{n-2}>2 \times 3^{n-3/2}$ (see Appendix~\ref{app:diew})~\cite{remark}.

We now illustrate how this DIEW can be used to detect genuine multipartite entanglement. For this, let us consider the noisy GHZ state $\rho=V|GHZ\rangle_n \,_n\!\langle GHZ|+(1-V)\openone/2^n$
characterized by the visibility $V$. Carrying out the measurements 
{$\cos((\tfrac{x_i-1}{3}-\tfrac{1}{6n})\pi)\,\sigma_x+\sin((\tfrac{x_i-1}{3}-\tfrac{1}{6n})\pi)\,\sigma_y$} on all parties, we obtain $I_n = 3^{n-1/2}V$, which violates (\ref{eq:witness}) provided that $V>2/3$. The DIEW \eqref{eq:witness} can thus detect in a DI way genuine $n$-partite entanglement in a noisy GHZ state for visibilities as low as $V=2/3$. This significantly improves over the threshold visibility $V=1/\sqrt{2}$ required to violate the DIEW based on the Mermin expression \eqref{mermin} or the different inequalities introduced in \cite{diew}. Note that in a DI setting it is not possible to detect the genuine multipartite entanglement in tripartite GHZ states below $V=1/2$ using projective measurements. Indeed, we have shown in this case that there exists a biseparable model reproducing all GHZ correlations (see Appendix~\ref{app:model}). 

In the case $n=3$, the DIEW (\ref{eq:witness}) takes the form  $I_3=E_3^3+E_3^4-E_3^6-E_3^7+E_3^9\leq 6\sqrt3$. It therefore involves only 18 expectation values,
compared to $27$ for a full tomography of a three-qubit system.
Let us stress, however, that contrary to usual entanglement witnesses $I_3$ is not restricted to two-dimensional Hilbert spaces, even though it uses observables with binary outcomes. 
For instance, if all parties perform the measurements $2\ketbra{\phi(x_i)}{\phi(x_i)}-\openone$ with $\ket{\phi(x_i)}=\frac1{\sqrt2}\left(\ket{0}+e^{i(6x_i-7)\pi/18}\ket{1}\right)$ on the three-qutrit state $\frac1{\sqrt3}\left(\ket{000} + \ket{111} + \ket{222}\right)$, then $I_3=6\sqrt3+8/3$, showing that the state is genuinely tripartite-entangled.


\textbf{3. General characterization of biseparable quantum correlations in the case $n=3$.}
Though the DIEW (\ref{eq:witness}) seems particularly well adapted to GHZ states, we cannot expect a single nor a finite set of DIEW to completely characterize the biseparable region, as illustrated in Fig.~1. It is thus desirable to derive a general method to decide whether arbitrary correlations are biseparable. Here we show how the semi-definite programming (SDP) techniques introduced in \cite{Navascues07,hierarchy} can be used to certify that the correlations observed in an experiment are genuinely tripartite entangled. 

Our approach is based on the observation that the tensor product separation $\rho_{AB}\otimes \rho_C$ at the level of states in the definition (\ref{stbis}) of biseparable states can be replaced by a commutation relation at the level of operators. Specifically, let $s=\{AB/C,AC/B,BC/A\}$ denotes the three possible partitions of the parties into two groups. Then, $P(abc|xyz)$ are biseparable quantum correlations if and only if there exist three arbitrary (not necessarily biseparable) states $\rho^s$ and three sets of measurement operators  $\{M_{a|x}^s,M_{b|y}^s,M_{c|z}^s\}$ such that
\be\label{eq:decomp}
P(abc|xyz)=\sum_s \text{tr}[M^s_{a|x}\otimes M^s_{b|y}\otimes M^s_{c|z}\rho^s]\,,
\ee
where measurement operators corresponding to an isolated party commute, i.e., $[M_{c|z}^{AB/C},M_{c'|z'}^{AB/C}]=0$, and similarly for the other partitions. The equivalence between (\ref{eq:corrs}) and (\ref{eq:decomp}) is established in Appendix~\ref{app:sdp}.
The problem of determining whether given correlations $P(abc|xyz)$ are biseparable thus amounts to finding a set of operators satisfying a finite number of algebraic relations (the projection defining relations of the type $M^s_{a|x}M^s_{a'|x}=\delta_{a,a'}M^s_{a|x}$, $\sum_{a}M^s_{a|x}=\openone$ and the commutation relations mentioned above) such that (\ref{eq:decomp}) holds. Such a problem is a typical instance of the SDP approach introduced in \cite{Navascues07, hierarchy} (see details in Appendix~\ref{app:sdp}). Specifically, it follows from the results of \cite{hierarchy} that one can define an infinite hierarchy of criteria that are necessarily satisfied by any correlations of the form (\ref{eq:decomp}) and which can be tested using SDP. If given correlations do not satisfy one of these criteria, we can conclude that they reveal genuinely tripartite entanglement. Further, it is possible  in this case to derive an associated DIEW from the solution of the dual SDP. Modulo a technical assumption, it can be shown that the hierarchy of SDP criteria is complete, that is, if given correlations are not biseparable this will necessarily show-up at some finite step in the hierarchy. 

\begin{table}
\begin{tabular}{ccc}
State & \qquad $V_{\min}$ with two settings & \qquad $V_{\min}$ with three settings \\
\hline
GHZ & $0.7071\simeq 1/\sqrt{2}$ & $0.6667\simeq 2/3$ \\
\hline
W & $0.7500\simeq 3/4$ & $0.7158$
\end{tabular}
\caption{Summary of numerical investigations.}\label{tab:tab}
\end{table}
\textbf{4. Application to GHZ and W states.}
Using finite levels of this hierarchy and optimizing over the possible measurements, we investigated the minimal visibilities above which the GHZ state $|GHZ\rangle$ and the W state $\ket{W}=(\ket{001}+\ket{010}+\ket{100})/\sqrt{3}$ exhibit correlations that are not biseparable (and thus reveal genuine tripartite entanglement) in the case of two and three measurement settings per party. Our results are summarized in Table~\ref{tab:tab}. For GHZ states, the reported visibility $V_\text{min}=2/3$ for three measurements per party correspond to the threshold required to violate the DIEW (\ref{eq:witness}), suggesting that this DIEW is optimal in this case. 
In the case of two measurements per party, we could not lower the visibility below the threshold $V_\text{min}=1/\sqrt{2}$, which corresponds to the visibility required to violate the DIEW based on Mermin expression (\ref{mermin}) and the DIEWs introduced in \cite{seevinck,diew}. Note, however, that for $V>1/\sqrt{2}$ the GHZ state violates Svetlichny's inequality \cite{svet} and thus exhibits genuine tripartite nonlocality. Thus for GHZ state the DIEWs introduced in \cite{seevinck,diew} do not improve over what can already be concluded using the standard notion of tripartite non-locality. On the other hand, our numerical explorations suggest that the visibilities $V_\text{min}=2/3$ for GHZ states with three measurements and $V_{\min}=3/4$ for W states with two measurements cannot be attained using the notion of genuine tripartite non-locality, illustrating the interest of the weaker notion of DIEW.

\textbf{Dicussion.}
To conclude, we comment on some possible directions for future research. First of all, note that by identifying the measurement settings ``$X_i$'' with $x_i=1$ and ``$Y_i$'' with $x_i=2$, the two-setting DIEW based on Mermin expression (\ref{mermin}) can be written as $E_3^3-E_3^5\leq 2\sqrt{2}$, which is of the same general form as the three-setting DIEW (\ref{eq:witness}). This suggests that the DIEWs based on (\ref{mermin}) and (\ref{eq:witness}) actually form part of a larger family of $m$-settings DIEWs. This question deserves further investigation. A second problem is to derive simple DIEWs that are adapted to W states and that can in particular reproduce the threshold visibilities obtained in Table~1. Finally, we have shown a practical method to characterize three-partite biseparable correlations using SDP. It would be interesting to understand how to generalize these results to the $n$-partite case. A possibility would be to combine the approach of \cite{Navascues07,hierarchy} with the symmetric extensions introduced in \cite{terhal}. This question will be investigated elsewhere. 

\paragraph*{Acknowledgments.}
This work was supported by the Swiss NCCRs QP and QSIT, the European ERC-AG QORE, and the Brussels-Capital region through a BB2B grant.


\appendix
\section{Biseparable quantum correlations}\label{app:def}
Here, we show that the definitions (\ref{eq:corrs}) and (\ref{eq:svetq}) of biseparable correlations are equivalent. For the sake of generality, we prove this equivalence in the $n$-partite scenario.
Let $t$ denote a subset of $\{1,\ldots,n\}$ containing from one to $n-1$ elements, and let $t'=\{1,\ldots,n\}\setminus t$ denote the complementary subset. The pair $(t,t')$ thus represent a partition of the $n$ parties into two non-empty groups. Let 
$\rho_{t}\otimes \rho_{t'}$ be an (unormalized) $n$-partite state that is product across the partition $(t,t')$. An $n$-partite state $\rho_{bs}$ is then biseparable if it can be written as a mixture
\begin{equation}\label{abs}
\rho_{bs}=\sum_{t}\sum_{k_t}  \rho^{k_t}_{t}\otimes \rho^{k_t}_{t'}\,.
\end{equation}
A state that cannot be written in the above form is genuinely $n$-partite entangled.

Let $P(a_1\ldots a_n|x_1\ldots x_n)=P(\bar{a}|\bar{x})$ be an $n$-partite probability distribution characterizing a Bell experiment with measurement settings $\bar{x}=(x_1\ldots x_n)$ and measurement outcomes $\bar{a}=(a_1\ldots a_n)$. We say that the correlations described by $P(\bar{a}|\bar{x})$ are biseparable quantum correlations iff
\begin{equation}\label{acorr}
P(\bar{a}|\bar{x})=\text{tr}\left[M_{a_1|x_1}\otimes\ldots\otimes M_{a_n|x_n}\,\rho_{bs}\right]
\end{equation}
for some biseparable state $\rho_{bs}$ and measurement operators $M_{a_i|x_i}$. 
Equivalently, as we will see, biseparable quantum correlations can be defined as those that can be written in the form
\be\label{aeq:svetq}
P(\bar{a}|\bar{x}) = 
\sum_{t}\sum_{k_{t}} P^{k_t}_Q(\bar{a}_t|\bar{x}_t)P_Q^{k_t}(\bar{a}_{t'}|\bar{x}_{t'})\,,
\ee
where we write $\bar{a}_t$ and $\bar{x}_t$ for the outcome and measurement settings of the parties belonging to the subset $t$ and where $P_Q^k(\bar{a}_t|\bar{x}_t)=\text{tr}\left[\bigotimes_{i\in t} M^k_{a_i|x_i}\,\rho^k_t\right]$ are arbitrary (unnormalized) quantum correlations for the parties in $t$. 

That any correlations of the form (\ref{acorr}) is of the form (\ref{aeq:svetq}) is immediate using the definition (\ref{abs}) of biseparable states.  Conversely, (\ref{aeq:svetq}) can be rewritten in the form
\be\label{aeq:svetq2}
P(\bar{a}|\bar{x}) = 
\sum_{t}\sum_{k_{t}} \text{tr}\left[\bigotimes_{i\in n} M^{k_t}_{a_i|x_i}\,\rho^{k_t}_t\otimes \rho^{k_t}_{t'}\right]\,.
\ee
This gives a representation that is almost of the form (\ref{acorr}), except that a different set of measurement operators corresponds to each term in the decomposition (\ref{abs}) of the biseparable state $\rho_{{bs}}$. This can be fixed by introducing local ancillas on each system acting as labels that indicate which term in the decomposition is considered: defining ${\rho_{\mathrm{bs}}	}=\sum_{t}\sum_{k_t} (\ketbra{t,k_t}{ t,k_t})_1\otimes\ldots\otimes(\ketbra{t,k_t}{ t,k_t})_n\otimes\rho^{k_t}_t\otimes\rho^{k_t}_{
t'}$, $M_{a_i|x_i}=\sum_{t}\sum_{k_t} (\ketbra{t,k_t}{ t,k_t})_i\otimes M^{k_t}_{a_i|x_i}$  we finally get $P(\bar{a}|\bar{x})=\text{tr}\left[M_{a_1|x_1}\otimes\ldots\otimes M_{a_n|x_n}\,\rho_{bs}\right]$, and thus any correlations of the form \eqref{aeq:svetq2} can be written as in Eq.~\eqref{acorr}.

Let $Q_t$ denote the set of quantum correlations defined for the parties belonging to $t$ and let $Q_{bs}$ denote the set of biseparable quantum correlations, which is a subset of the $n$-partite quantum correlations. 
From (\ref{aeq:svetq}), it is clear that $Q_{bs}$ is convex and that its extreme points are of the form $P_{Q}^{ext}(\bar{a}_t|\bar{x}_t)P^{ext}_Q(\bar{a}_{t'}|\bar{x}_{t'})$ where $P_Q^{ext}(\bar{a}_t|\bar{x}_t)$ is an extremal point of the set $Q_t$.

\section{DIEWs for genuine $n$-partite entanglement}\label{app:diew}

Here we derive the maximum value that the expression $I_n$, see Eq.~\eqref{eq:witness}, can take for measurements made on a biseparable quantum state (the biseparable bound), and for Svetlichny models (the Svetlichny bound). From (\ref{aeq:svetq}), it follows that the biseparable bound is obtained by maximizing $I_n$ over all correlations of the form
\be\label{beq:svetq}
P(\bar{a}|\bar{x}) = 
\sum_{t}\sum_{k_{t}} P^{k_t}_Q(\bar{a}_t|\bar{x}_t)P_Q^{k_t}(\bar{a}_{t'}|\bar{x}_{t'})\,.
\ee
In a Svetlichny model, on the other hand, one considers correlations of the form 
\be\label{beq:svets}
P(\bar{a}|\bar{x}) = 
\sum_{t}\sum_{k_{t}} P^{k_t}(\bar{a}_t|\bar{x}_t)P^{k_t}(\bar{a}_{t'}|\bar{x}_{t'})\,,
\ee
i.e. no constraints (apart from positivity and normalization) are imposed on the joint terms $P^{k_t}(\bar{a}_t|\bar{x}_t)$. But more refined models \`a la Svetlichny can be introduced, see e.g. \cite{pirbancsca}, the more constraining one being the one where the joint terms $P^{k_t}(\bar{a}_t|\bar{x}_t)$ are assumed to be no-signalling. We will compute the bound on $I_n$ assuming these no-signalling constraints and see that even in this case there exists a gap between the quantum biseparable and Svetlichny bounds (note also that the no-signalling bound that we will derive actually coincides with the more general unconstrained Svetlichny bound because $I_n$ is an inequality involving only full $n$-partite correlators).

We will now prove the biseparable bound given in Eq.~\eqref{eq:witness} for $n\geq 3$ using induction on the number of parties $n$. 
Let us start with the first step of the induction, which consists of showing that the inequality holds for $n=3$. By linearity of $I_3$ in the probabilities, convexity of the decomposition \eqref{eq:svetq} for tripartite biseparable quantum correlations and the fact that $I_3$ is invariant under any permutation of the parties, it is sufficient to prove the bound for correlations of the form $P_Q(ab|xy)P_Q(c|z)$. Moreover, it is sufficient to consider the case where $P_Q(c|z)$ is extremal, i.e., one where every $c$ is determined unambiguously as a function of $z$. To this end, let us label the eight distinct deterministic assignments of outcome $c$ for given input $z$ by $\gamma=(\gamma_1,\gamma_2,\gamma_3)\in\{\pm1\}^3$, where the deterministic (quantum) probability distributions are such that  $P_Q(c|z)=P_\gamma(c|z)=1$ if $\gamma_z=c$ and $P_\gamma(c|z)=0$ otherwise.
Substituting these eight possible strategies into $I_3$ [see \eqref{eq:witness}] gives eight different bipartite Bell expressions for the system $AB$:
\begin{equation}\label{Eq:I3gamma}
I_{3,\gamma}=\sum_{k=2}^{6}g_\gamma(k)E_2^k
\end{equation}
where $g_\gamma(k)=\sum_{z=1}^3 \gamma_zf_{k+z-3}$. Determining the biseparable bound of $I_3$ therefore amounts to finding the maximum of the above Bell expressions over arbitrary bipartite quantum correlations $P_Q(ab|xy)=\text{tr}[M_{a|x}\otimes M_{b|y} \rho]$, i.e., finding the corresponding Tsirelson bounds. 
It can be easily verified that the eight expressions $I_{3,\gamma}$ are either vanishing or equivalent, under relabelling of inputs and outputs, to two times the 3-input chained Bell inequality~\cite{Braunstein90}.  The biseparable bound on $I_3$ is thus equal to the two times the Tsirelson bound of the 3-input chained Bell inequality~\cite{Wehner06}, i.e., $I_3\leq6\sqrt3$, which is in agreement with Eq.~\eqref{eq:witness}. 

Next, we need to show that whenever inequality \eqref{eq:witness} holds for $n$ parties, it must also hold for $(n+1)$ parties. By linearity of $I_n$ in the probabilities and convexity of the decomposition (\ref{beq:svetq}), it is sufficient to compute the bound over all product correlations of the form $P_Q(\bar{a}_t|\bar{x}_t)P_Q(\bar{a}_{\hat{t}}|\bar{x}_{\hat{t}})$. Moreover, since $I_{n+1}$ is symmetric under any permutation of parties, it is sufficient to consider the biseparations $t=\{1,2,\ldots,k\}$, $t'=\{k+1,\ldots,n+1\}$ where $1\leq k \leq \lfloor \frac{n}2 \rfloor$. With obvious notation, we write the corresponding correlations as $P_Q(a_{1,\ldots,k}|x_{1,\ldots,k})P_Q(a_{k+1,\ldots,n+1}|x_{k+1,\ldots,n+1})$. 
Bayes' rule and the fact that quantum correlations satisfy the no-signaling principle allow us to write the quantum correlations for parties $k+1$ to $n+1$ as $P_Q(a_{k+1,\ldots,n}|x_{k+1,\ldots,n+1}, a_{n+1})P_Q(a_{n+1}|x_{k+1,\ldots,n+1})=P_Q(a_{k+1,\ldots,n}|x_{k+1,\ldots,n},  a_{n+1}, x_{n+1})P_Q(a_{n+1}|x_{n+1})$.
With straightforward algebra, it can be seen that the quantity $E_{n+1}^k=\sum_{\bar x}\delta(\sum_{i=1}^{n+1}x_i=k)E(\bar{x})$ for this type of distribution can be rewritten as
\begin{equation}\begin{split}
E_{n+1}^k
& =\sum_{x_{n+1}=1}^3 \sum_{a_{n+1}=-1}^1{a_{n+1}} P_Q(a_{n+1}|x_{n+1})\times \\
&\ \ \ \ \ \ \ \ \ \ \ \ \ \ \ \ \ \ \ \ \ \ \ E^{k-x_{n+1}}_{n}(a_{n+1},x_{n+1})
\end{split}\end{equation}
where the $n$-partite quantity $E^{k-x_{n+1}}_{n}(a_{n+1},x_{n+1})$ depends on the values of $a_{n+1}$ and $x_{n+1}$ since it is evaluated on the distribution $P_Q(a_{1,\ldots,k}|x_{1,\ldots,k})P_Q(a_{k+1,\ldots,n}|x_{k+1,\ldots,n},  a_{n+1},x_{n+1})$. Inserting this expression in the definition $I_{n+1}=\sum_{k=n+1}^{3(n+1)} f_{k-n} E_{n+1}^k$, we obtain
\begin{equation}\label{b5}
I_{n+1}=\sum_{x_{n+1}=1}^3 \sum_{a_{n+1}=-1}^1 a_{n+1} P_Q(a_{n+1}|x_{n+1}) I_{n}^{x_{n+1}}(a_{n+1},x_{n+1})
\end{equation}
where we have defined $I_n^{j}=\sum_{k=n+1}^{3n+3} f_{k-n} E^{k-j}_{n}=\sum_{k=n}^{3n} f_{k-n+j}E^k_n$ (note that $E^k_n$ is different from zero only if $n\leq k\leq 3n$) and write $I_n^j(a_{n+1},x_{n+1})$ to remind that $I_n^j$ is evaluated on the distribution $P_Q(a_{1,\ldots,k}|x_{1,\ldots,k})P_Q(a_{k+1,\ldots,n}|x_{k+1,\ldots,n},  a_{n+1},x_{n+1})$ which depends on  $a_{n+1}$ and $x_{n+1}$. Using the fact that $a_{n+1}=\pm 1$ and that the probabilities $P(a_{n+1}|x_{n+1})$ are bounded between $0$ and $1$, it follows from (\ref{b5}) that
\begin{equation}\label{b6}
I_{n+1}\leq \sum_{x_{n+1}=1}^3 |I_{n}^{x_{n+1}}(a_{n+1},x_{n+1})|\,.
\end{equation}
Let $\max |I_n^j|=\max \pm I_n^j$ denote the maximal value of the quantity $\pm I_n^j$ taken by any 
biseparable correlations. Since the distribution $P_Q(a_{1,\ldots,k}|x_{1,\ldots,k})P_Q(a_{k+1,\ldots,n}|x_{k+1,\ldots,n},  a_{n+1}x_{n+1})$ is biseparable, we clearly have that $|I_{n}^{x_{n-1}}(a_{n+1},x_{n+1})|\leq \max \pm I_n^j$, independently of the values of $a_{n+1}$ and $x_{n+1}$. We thus find
\begin{equation}\label{b7}
I_{n+1}\leq \sum_{j=1}^3 \max \pm I_{n}^{j}\,.
\end{equation}

An important point to note now is that $\pm I_n^j$  for all $j=1,2,3$ are equivalent to $I_n$, i.e., one can obtain the expression for $I_n$ by starting from any of $\pm I_n^j$ and applying a different labeling for the inputs and/or outputs. To see that this is the case, we first note from the definition of $f_k$ that $-I_n^3=I_n$. In other words, we can obtain $I_n^3$ by starting from $I_n$ and applying the mapping $a_n\to -a_n$ to each of the $E_n^k$. Clearly, the same argument also demonstrates the equivalence between $I^j_n$ and $-I^j_n$. Next, note that if the following mappings are applied to the definition of $E_n^k$, namely $x_n\to x_n'=(x_n \mod 3)+1$ and $a_n\to a_n'=a_n$ except for $x_n=3$ in which case $a_n' = -a_n$, then we can also obtain the expression for $I_n^j$ from $I_n^{j+1}$, thus showing their equivalence. We thus have that $\max \pm I_n^{j}\leq \max I_n$ and thus
\begin{equation}\label{b8}
I_{n+1}\leq 3 \max  I_{n}\,.
\end{equation}
Now, by the induction hypothesis we know that $\max I_n\le 2\times 3^{n-3/2}$ for all biseparable correlations . It then follows from Eq.~(\ref{b8}) that $I_{n+1}\le 2\times 3^{n-1/2}$, which completes the proof.

Note that, following the above reasoning, any witness for $n$ parties and $m$ inputs per party which can be written in the form $\sum_k f_k E_n^k$, where $f_k:\mathbb{Z}\to\mathbb{R}$ is a function satisfying $f_{k+m} = \pm f_{k}$, can be generalized to more parties.

Note also that the above biseparable bound for the witness \eqref{eq:witness} is tight, as it can always be achieved by performing the following local measurements 
\begin{equation}
\left\{
\begin{array}{c@{\quad:~ }l@{\quad}}
\cos\phi(x_i)\sigma_x+\sin\phi(x_i)\sigma_y & i=1,\ldots,n-1 \\
\sigma_z &  i=n  
 \end{array}\right. ,
\end{equation}
with $\phi(x_i)=(\tfrac{x-1}{3} +\tfrac{1}{6(n-1)})\pi$ on the biseparable state 
\begin{equation}
\ket{\psi_n}=\ket{GHZ_{n-1}}\otimes\ket{0}.
\end{equation}

Finally, let us note that the same procedure as the one detailed above can be followed to obtain the (no-signalling) Svetlichny bound of $I_{n+1}$. The only difference is that in the first step of the induction proof, we must compute the no-signalling bound of the $I_3$ inequality instead of the quantum biseparable bound. As in the quantum biseparable case, this reduces to computing twice the (bipartite) no-signalling bound of the 3-input chained Bell inequality, which gives $I_3\leq2\times 6=12$  and thus $I_{n}\leq 4\times3^{n-2}$. To show that this bound is tight consider a strategy where the $n$-th party is separated from the rest and always outputs 1. Using an analysis similar to the one leading to Eq.~\eqref{b6}, we can then see that the remaining $(n-1)$ parties are playing an effective game defined by $I_{n-1}^1+I_{n-1}^2+I_{n-1}^3$ which can be shown to be equivalent to twice the expression $I_{n-1}$. It is a simple exercise to show that the algebraic maximum of $I_{n-1}$ is $2\times 3^{n-2}$ and that this is always achievable by $(n-1)$ players that are constrained only by the no-signalling principle. Therefore, with this particular strategy, one achieves $4\times3^{n-2}$, which saturates the  bound derived above.

\section{Biseparable model for projective measurements on the tripartite GHZ state}\label{app:model}
Here we present a biseparable model that simulates von Neumann measurements on noisy tripartite GHZ states
\be\label{eq:state}
\rho=V\frac{\ketbra{000+111}{000+111}}2+(1-V)\frac{\openone}8
\ee
of visibility $V=\frac12$. Clearly this allows to simulate states with $V<\frac12$ as well, by mixing this model with another one in which all parties produce uniformly random outcomes.

The model is presented as a protocol in which a source distributes quantum states and random variables to the parties. All parties can share the random variables, but since only biseparable states may be used in the model, not more than two parties at a time can share a quantum state. After distribution, the parties receive their respective measurement directions $\vec x$, $\vec y$ or $\vec z$ belonging to the Bloch sphere, and measure their quantum system and/or process the information they received accordingly in order to produce binary outcomes $A,B,C=\pm1$.

We check that the outcomes produced by the model are identical to the ones found when measuring state \eqref{eq:state} in the given bases.

\subsection{The model}
Before the parties receive their inputs, a common source chooses a party $p$ at random, say Charlie ($p=\text{C}$), and sends him the vector
\be
\vec\lambda=(\sin\alpha\cos\beta,\sin\alpha\sin\beta,\cos\alpha),
\ee
uniformly chosen on the sphere $S^2$. The source then also provides the two other parties, Alice and Bob in this case, with the quantum state
\be\label{eq:phiab}
\ket{\Phi_{AB}}=\cos\frac{\alpha}2\ket{00} + \sin\frac{\alpha}2 e^{-i\beta}\ket{11}.
\ee
Moreover, the source sends to all parties the signs $s_{AB},s_{AC},s_{BC}=\pm1$, which are independently and identically distributed with $\Prob(s=+1)=\frac{2+\sqrt3}4$.

At the time of measurement, Alice and Bob measure their system according to $\vec x$ and $\vec y$ and get outcomes $a,b=\pm1$, while Charlie calculates $c=\sg(\vec\lambda\cdot\vec z)$. The parties then output respectively $A=s_{AB}s_{AC}a$, $B=s_{AB}s_{BC}b$ and $C=s_{BC}s_{AC}c$.

\subsection{Correlations obtained by the model}
Here we compare the correlations that are created when Alice, Bob and Charlie apply the biseparable model above to the ones that they would get by measuring state~\eqref{eq:state}. Since the model treats equally each party, its correlations are symmetric under exchange of the parties and we only need to check the following three relations:
\begin{align}
\langle A \rangle &= 0\ \label{eq:corrs1}\\
\langle AB \rangle &= \frac12 \cos\theta_x \cos\theta_y\ \label{eq:corrs2}\\ 
\langle ABC \rangle &= \frac12 \sin\theta_x \sin\theta_y \sin\theta_z \cos(\varphi_x+\varphi_y+\varphi_z). \label{eq:corrs3}
\end{align}
Here we parametrized the parties' measurements in terms of spherical coordinates:
\be
\vec x=(\sin\theta_x\cos\varphi_x,\sin\theta_x\sin\varphi_x,\cos\theta_x)
\ee
and similarly for $\vec y$ and $\vec z$.

For definiteness, in the following we use the brackets $\langle q\rangle$ to express the expectation value of a quantity $q$ with respect to a quantum state, and the bar
\be
\overline{f(\alpha,\beta)}=\frac1{4\pi}\int_0^{2\pi} d\beta \int_0^\pi d\alpha \sin\alpha f(\alpha,\beta)
\ee
for the average of a function $f(\alpha,\beta)$ over the random variable $\alpha$ and $\beta$. We always start by considering that Charlie is the special party chosen by the source, to whom the hidden vector $\vec\lambda$ is sent, and perform the symetrization afterwards. Symmetrized quantities $q_\circlearrowleft=\frac13(q_{p=\text{A}}+q_{p=\text{B}}+q_{p=\text{C}})$ are indexed by the symbol $\circlearrowleft$.

For simplicity we average over the depolarization signs $s_{AB},s_{AC},s_{BC}$ only at the end of the calculation.

\bigskip

\paragraph{Single-party expectation value}
One can easily verify that the outcomes of each party is locally random according to the model, in agreement with equation \eqref{eq:corrs1}.

\bigskip

\paragraph{Two-party correlators}
Let us first calculate correlations $\overline{\langle ab\rangle}$, $\overline{\langle ac\rangle}$ and $\overline{\langle bc\rangle}$ for the case in which Charlie receives the vector $\vec\lambda$, and average over the choice of the party $p$ being alone afterwards.

The first term is easily found from equation \eqref{eq:phiab}, which gives
\be\label{eq:ab}
\langle ab \rangle = \cos\theta_x \cos\theta_y + \sin\alpha\sin\theta_x\sin\theta_y\cos(\beta+\varphi_x+\varphi_y)
\ee
and thus $\overline{\langle ab \rangle}=\cos\theta_x\cos\theta_y$.

\bigskip
To compute the correlation $\overline{\langle ac \rangle}$, it is useful to write Alice's state as
\be
\rho_A=\text{tr}_B(\ketbra{\Phi_{AB}}{\Phi_{AB}})=\cos^2\frac{\alpha}2\ketbra{0}{0}+\sin^2\frac{\alpha}2\ketbra{1}{1}.
\ee
The expectation value for Alice's outcome is then $\langle a \rangle=\cos\alpha\cos\theta_x$.

Concerning Charlie, his outcome $c$ is totally determined by $\vec \lambda$ and $\vec z$. For simplicity we assume that $0\leq\theta_z\leq\pi/2$, but the other situation can be treated similarly. In this case, for $\alpha\leq\frac\pi2-\theta_z$, Charles always has $c=+1$, and for $\alpha\geq\frac\pi2+\theta_z$, he always has $c=-1$. Now for $\alpha\in[\frac\pi2-\theta_z,\frac\pi2+\theta_z]$, we can write $\mathcal{I}_+=[-\Phi_c+\varphi_z,\Phi_c+\varphi_z]$ the interval of $\beta$ for which the product $\vec\lambda\cdot\vec z \geq 0$, and $\mathcal{I}_-=[\varphi_z-\pi,\varphi_z+\pi]\backslash \mathcal{I}_+$ its complement, with $\Phi_c=\arccos(-\cot\alpha\cot\theta_z)$. We thus have the three following cases:

\begin{enumerate}
 \item $\alpha\leq\frac{\pi}2-\theta_z$. In this case $c=+1$, which gives $\langle ac\rangle=\cos\alpha\cos\theta_x$.
 \item $\alpha\geq\frac{\pi}2+\theta_z$. In this case $c=-1$, which gives $\langle ac\rangle=-\cos\alpha\cos\theta_x$.
 \item $\frac{\pi}2-\theta_z\leq\alpha\leq\frac{\pi}2+\theta_z$. In this case one has $\langle ac\rangle=\cos\alpha\cos\theta_x(\chi_\mathcal{I_+}(\beta)-\chi_\mathcal{I_-}(\beta))$ where \begin{equation}
\chi_\mathcal{I}(\beta)=\begin{cases}1&\text{if }\beta\in\mathcal{I}\\0&\text{if }\beta\notin\mathcal{I}\end{cases}
\end{equation}
is the indicator function.
\end{enumerate}

In total, after integration over $\alpha$ and $\beta$, this gives $\overline{\langle ac\rangle}=\frac12\cos\theta_x\cos\theta_z$.

\bigskip

Similarly one can check that $\overline{\langle bc \rangle} = \frac12\cos\theta_y\cos\theta_z$, and so once averaged over the choice of the party being alone, the bipartite correlations are given by
\begin{equation}
\overline{\langle ab \rangle}_\circlearrowleft = \frac23\cos\theta_x\cos\theta_y.
\end{equation}

\bigskip

Finally, we need to apply the signs $s_{AB}$, $s_{AC}$, $s_{BC}$. Since the expectation value of the product of two signs is $\langle s_{AB} s_{AC}\rangle=\frac34$, it follows that overall the correlation between two outcomes produced by the model is
\be
\langle AB \rangle=\frac12\cos\theta_x\cos\theta_y,
\ee
in agreement with equation \eqref{eq:corrs2}.

\bigskip

\paragraph{Three-party correlators}
We now proceed to calculate the tripartite correlations that are created by the model. Let us consider the three preceding cases separately again:
\begin{enumerate}
 \item $\alpha\leq\frac{\pi}2-\theta_z$. In this case $c=+1$, so $\langle abc\rangle=\langle ab \rangle$ as given by equation \eqref{eq:ab}.
 \item $\alpha\geq\frac{\pi}2+\theta_z$. In this case $c=-1$, so $\langle abc\rangle=-\langle ab \rangle$.
 \item $\frac{\pi}2-\theta_z\leq\alpha\leq\frac{\pi}2+\theta_z$. In this case one has $\langle abc\rangle=\langle ab \rangle(\chi_\mathcal{I_+}(\beta)-\chi_\mathcal{I_-}(\beta))$.
\end{enumerate}

In total, the tripartite correlation are found after integration over $\alpha$ and $\beta$ to be
\be
\overline{\langle abc\rangle}=\frac12\sin\theta_x\sin\theta_y\sin\theta_z\cos(\varphi_x+\varphi_y+\varphi_z),
\ee
which does not depend on which party $p$ is alone.

One can check that the application of the signs $s_{ij}$ has no influence on the tripartite correlations since they cancel out, and so $\langle ABC\rangle=\overline{\langle abc\rangle}$. The model thus reproduces the expected correlations \eqref{eq:corrs3}.

\section{Characterizing biseparable correlations through SDP in the tripartite case ($n=3$)}\label{app:sdp}
It has been shown in \cite{Navascues07} how to define a hierarchy of SDP that characterizes the set $Q_2$ of bipartite quantum correlations. Note that biseparable quantum correlations $P(abc|xyz)$ can be written as a finite sum of such bipartite quantum correlations using the fact that each $P^k(c|z)$ in (\ref{eq:svetq}) can be taken to be extremal and using the fact that there are a finite number of such extremal points corresponding to classical, deterministic strategies. It therefore follows that biseparable quantum correlations can be characterized using a finite number of the SDP hierarchies introduced in \cite{Navascues07}. Note however that the number of single-party deterministic strategies, and thus the number of terms in Eq.~\eqref{eq:svetq}, grows exponentially with the number of measurement settings, making such an approach impractical even for small problems. Here we introduce an alternative SDP approach that has better scaling properties. 

Our approach is based on the observation that the tensor product separation $\rho_{AB}\otimes \rho_C$ at the level of states can be replaced by a commutation relation at the level of operators. Specifically, let $s=\{AB/C,AC/B,BC/A\}$ denote the three possible partitions of the parties into two groups. Then, $P(abc|xyz)$ are biseparable quantum correlations if and only if there exist three arbitrary (not necessarily biseparable) states $\rho^s$ and three sets of measurement operators $\{M_{a|x}^s,M_{b|y}^s,M_{c|z}^s\}$ such that (\ref{eq:decomp}) holds, i.e. 
\begin{subequations}\label{aeq:decomp}
\be\label{aeq:decomp1}
P(abc|xyz)=\sum_s \text{tr}[M^s_{a|x}\otimes M^s_{b|y}\otimes M^s_{c|z}\rho^s]\,,
\ee
where measurement operators corresponding to an isolated party commute, i.e.,
\be
\begin{split}\label{aeq:decomp2}
[M_{c|z}^{AB/C},M_{c'|z'}^{AB/C}]&=0\\
[M_{b|y}^{AC/B},M_{b'|y'}^{AC/B}]&=0\\
[M_{a|x}^{BC/A},M_{a'|x'}^{BC/A}]&=0.
\end{split}
\ee
\end{subequations}
Here, for convenience, the normalization factor of the RHS of \eqref{aeq:decomp1} has been absorbed in the quantum states $\rho^s$.

Let us start by showing that any correlations of the form \eqref{eq:svetq} is of the form \eqref{aeq:decomp}. Note first the well-known fact that if the dimension of the Hilbert space is not fixed, any single-party correlations $P^k(c|z)$ admits a quantum representation $P^k(c|z)=\text{tr}[M^k_{c|z}\,\rho_C^k]$ where the operators $M^k_{c|z}$ commute among themselves. Indeed, let $\overline{c}=(c_1,\ldots,c_m)$ and $P(\overline{c})=\prod_{z=1}^m P(c_z|z)$, then a quantum representation of $P^k(c|z)$ is achieved by defining $\rho_C^k=\sum_{\overline{c}}P(\overline{c})|\overline{c}\rangle\langle\overline{c}|$ and $M^k_{c|z}=\sum_{\overline{c}} \delta(c_z,c) |\overline{c}\rangle\langle\overline{c}|$. We can thus write in Eq.~(\ref{eq:svetq}), e.g., $\sum_k P^k_Q(ab|xy)P^k(c|z)=\text{tr}[M_{a|x}\otimes M_{b|y}\otimes M_{c|z}(\rho_{AB}\otimes\rho_C)]$ with the operators $M_{c|z}$ commuting between themselves. The decomposition (\ref{eq:svetq}) is thus clearly a particular case of (\ref{aeq:decomp}). 

Let us now show the converse. Let us suppose first that $s=AB/C$ and let $\overline{c}=(c_1,\ldots,c_m)$. Since the projectors $M^s_{c|z}$ commute, the projectors $M(\overline{c})=\prod_{z=1}^m M^s_{c_z|z}$ defines a valid measurement and thus $\rho_{\overline{c}}=\text{tr}_C[M^s_{\overline{c}}\rho^s M^s_{\overline{c}}]$ is a proper (unormalized) state of system $AB$. Define $P^{\overline{c}}_Q(ab|xy)=\text{tr}[M^s_{a|x}\otimes M^s_{b|y}\rho_{\overline{c}}]$ and $P^{\overline{c}}(c|z)$ as the deterministic point satisfying $P^{\overline{c}}(c|z)=\delta(c_z,c)$. It is then easy to see using the properties of the operators $M^s_{c|z}$ that $\text{tr}[M^s_{a|x}\otimes M^s_{b|y}\otimes M^s_{c|z} \rho^s]=\sum_{\overline{c}} \delta(c_z,c)\text{tr}[M_{a|x}\otimes M_{b|y}\otimes M_{\overline{c}} \,\rho^s]=\sum_{\overline{c}}\delta(c_z,c)\text{tr}[M_{a|x}\otimes M_{b|y} \rho_{\overline{c}}]=\sum_{\overline{c}}P^{\overline{c}}_Q(ab|xy)P^{\overline{c}}(c|z)$. This last expression is of the same form as the first series of terms in (\ref{eq:svetq}) (since any deterministic point $P^{\overline{c}}(c|z)$ can be seen as a single-party quantum point $P_Q(c|z)$). A similar argument for the other values of $s$ implies that any correlations of the form (\ref{aeq:decomp}) are of the form (\ref{eq:svetq}). 

We thus have reduced the problem of determining whether given correlations $P(abc|xyz)$ are biseparable to the problem of finding a set of operators satisfying a finite number of algebraic relations (the projection defining relations of the type $M^s_{a|x}M^s_{a'|x}=\delta_{a,a'}M^s_{a|x}$, $\sum_{a}M^s_{a|x}=\openone$ and the commutation relations \eqref{aeq:decomp2}) such that (\ref{aeq:decomp1}) holds. Such a problem is a typical instance of the SDP approach introduced in \cite{hierarchy}, which generalizes the results of \cite{Navascues07}. Specifically, it follows from the decomposition (\ref{aeq:decomp}) and the results of \cite{hierarchy} that it is possible to define an infinite hierarchy of sets $\{Q^j_{2/1}\,:\,j=1,\ldots,\infty\}$ with the following properties: $i)$ deciding whether given correlations belongs to $Q^j_{2/1}$ can be determined using SDP (though the size of the SDP increases with $j$); $ii)$ the sets $Q^j_{2/1}$ approximate better and better $Q_{2/1}$ from the outside, i.e.,  $Q_{2/1}\subseteq Q^j_{2/1}$ and $Q^{j+1}_{2/1}\subseteq Q^j_{2/1}$ for all $j$. If given correlations are found not to belong to some $Q_{2/1}^j$, we can thus conclude that they do not belong to $Q_{2/1}$ and that they reveal genuinely tripartite entanglement.  Further, it is possible  in this case to derive an associated DIEW from the solution of the dual SDP. Note that modulo the assumption that the tensor product between the three measurement operators in (\ref{aeq:decomp1}) is equivalent to the commutation of these operators --- an assumption that is true in finite-dimensional Hilbert space, but which still represents an open question in an infinite-dimensional Hilbert space --- it can be shown that the hierarchy of relaxations converges to the set of biseparable states, i.e., $\lim_{j\rightarrow\infty}Q^j_{2/1}=Q_{2/1}$. Thus if given correlations are not biseparable this will necessarily show-up at some finite step in the hierarchy.

\end{document}